\documentclass[twocolumn,showpacs,preprintnumbers,amsmath,amssymb,aps,prl,superscriptaddress]{revtex4-1}
\usepackage{color}
\usepackage{amsfonts,stackengine,graphicx}
\setstackgap{S}{-1.5pt}
\usepackage{textcomp} 
\usepackage{subeqnarray}
\usepackage[caption=false]{subfig} 
\usepackage{ulem}
\usepackage{color}

\usepackage{amsmath}

\begin{document}


\title{Beauty and structural complexity}

\author{Samy Lakhal}
\affiliation{Art in Research, 33 rue Censier, 75005 Paris, France}
\affiliation{Ladhyx UMR CNRS 7646, Ecole polytechnique,
91128 Palaiseau Cedex, France}
\affiliation{Chair of Econophysics \& Complex Systems, Ecole polytechnique, 91128 Palaiseau Cedex, France}
 \author{Alexandre Darmon}
\affiliation{Art in Research, 33 rue Censier, 75005 Paris, France}
 \author{Jean-Philippe Bouchaud}
 \affiliation{Chair of Econophysics \& Complex Systems, Ecole polytechnique, 91128 Palaiseau Cedex, France}
 \affiliation{Capital Fund Management, 23 rue de l'Universit\'e 75007 Paris, France \bigskip}
\author{Michael Benzaquen}%
 \email{michael.benzaquen@polytechnique.edu}
\affiliation{Ladhyx UMR CNRS 7646, Ecole polytechnique,
91128 Palaiseau Cedex, France}
\affiliation{Chair of Econophysics \& Complex Systems, Ecole polytechnique, 91128 Palaiseau Cedex, France}
 \affiliation{Capital Fund Management, 23 rue de l'Universit\'e 75007 Paris, France \bigskip}

\date{\today}

\begin{abstract}
We revisit the long-standing question of the relation between image appreciation and its statistical properties. We generate two different sets of random images well distributed along three measures of entropic complexity. We run a large-scale survey in which people are asked to sort the images by preference, which reveals maximum appreciation at intermediate entropic complexity. We show that the algorithmic complexity of the coarse-grained images, expected to capture structural complexity while abstracting from high frequency noise, is a good predictor of preferences. Our analysis suggests that there might exist some universal quantitative criteria for aesthetic judgement.

\end{abstract}

\maketitle

What makes a beautiful image? Is there such a thing as universal beauty? These puzzling yet fascinating questions have been tackled many times in the past within several disciplines, including philosophy, psychology, arts or mathematics \cite{fechner1876vorschule,eco2004storia, gombrich1979arte,vessel2010beauty, eysenck1941critical, bar2006humans, spehar2003universal, alvarez2016fractal,Hagerhall2004,mcmanus2005symmetry,graham2008statistical,Forsythe2011}. According to Kant, \textit{Is beautiful that which pleases universally without a concept} \cite{kant2000critique}. The idea of an intelligible beauty appeared in ancient Greece, where Nature was believed to be a \textit{cosmos} constituting a principle of order and harmony. The proportions between the constitutive elements of each being are rightfully defined, whether it is a work of art, a living organism or a city \cite{aujaleu1997sensibilite}. Following the Greeks, the Baroque and Renaissance artists also believed in a universal beauty, and it is striking that their arts partially rely on a mathematisation of the artistic representation (symmetry \cite{mcmanus2005symmetry}, proper geometric proportions as given by the \textit{golden number} \cite{fechner1876vorschule}, etc.). In other terms, the belief that there must be scientific grounds to the conception of what is artistic or beautiful has been out there for quite some time. Yet, the very idea of a universal beauty is a longstanding debate which has known many ruptures through the history of art~\cite{eco2004storia} and still opposes a number a great modern thinkers.

Physicists' interest in the subject is more recent. Stephens \textit{et al.} \cite{stephens2013statistical} showed that natural images were critical in the thermodynamic sense and proposed a theory for the \textit{Thermodynamics of Natural Images}. While, as pointed out above, many would consider \textit{quantitative aesthetics} to be an oxymoron, and indeed it would be rather nonsensical to aim at building a fully consistent theory of pictorial art, we, as physicists, believe there is some room for a quantitative analysis. For example one could easily argue that an aesthetically appealing image often results from a subtle balance between regularities and surprises. Indeed, it seems rather plausible to think that while one might find dull an image that is too regular (no surprises), one may also feel lost in front of an image with no recognisable shapes or structures to hang on to (too much surprise). As argued in~\cite{bouchaud2008models}, \textit{ Total chaos is disquieting. Too much regularity is boring. Aesthetics is perhaps the territory in-between}. Provided one agrees with such statements, these ideas clearly suggest that one could design an entropy-like function to quantify this subtle and complex equilibrium.

To address this question we run a large-scale survey in which people are asked to sort by preference two different sets of random images well distributed along three measures of entropic complexity: Fourier Magnitude's slope, fractal dimension and compression rate. The paper is organised as follows. We first present and motivate our image generation methods. We then present and analyse the results of the survey. Finally, we argue that algorithmic complexity of the coarse-grained images is a rather good proxy for image appreciation, and conclude.

There exist many possible measures of image complexity, relying on e.g. their mathematical properties \cite{Koch2010,Desolneux2008}, their physical properties \cite{stephens2013statistical,ImageRenormalization,ImageRenormalization2}, or even their cognitive impact \cite{Field1987,Spehar2015,Taylor2011}. Here we choose to work with three simple measures that can be easily computed unequivocally for any digital 2D image. The first one is the \textit{magnitude slope} $\alpha$   defined as the logarithmic slope of the radially averaged Fourier magnitude $\mathcal M(k) = \langle \hat u(k,\theta) \rangle_\theta $, where $\hat u(k,\theta)$ denotes the Fourier Transform of the image greyscale intensity $u(r,\phi)$, and $\mathcal M(k) \sim k^{\alpha}$. The second is the \textit{fractal dimension} $d_\mathrm{f}$ computed using the Minkowski-Bouligand box-counting method~\cite{dubuc1987variation}. After transforming the image to B\&W using an intensity threshold ensuring  two equally-populated levels \cite{stephens2013statistical}, the fractal dimension follows $N(\epsilon)\sim \epsilon^{-d_\mathrm{f}}$ where $N(\epsilon)$ is the number of boxes of size $\epsilon$ containing both black and white features. The third is the \textit{compression rate} or \textit{algorithmic complexity} $\tau$  computed as the ratio between the sizes of the PNG compressed and uncompressed image.

\begin{figure*}[t!]
\includegraphics[width=1\textwidth]{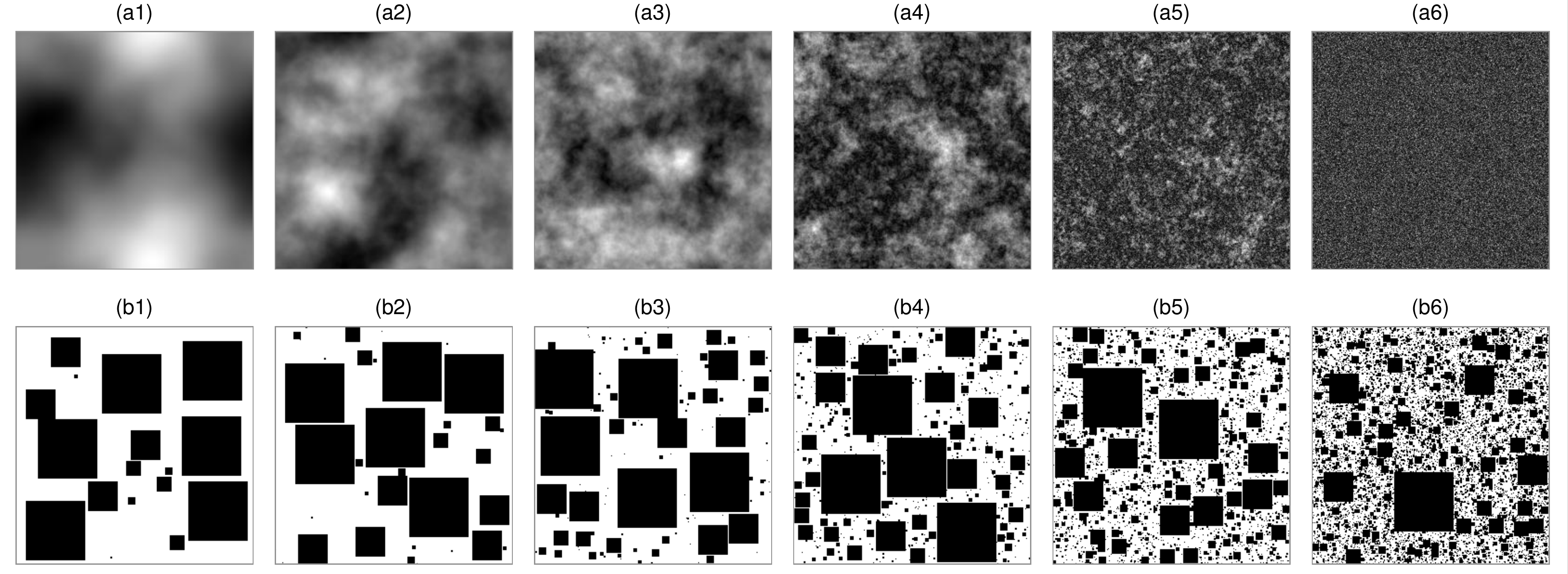}
\caption{\label{fig:images}(a) Fourier Magnitude-generated images, and (b) Box-counting-generated images, both series with increasing complexity from left to right. These images were used for our large-scale survey.}
\vspace{-0.2cm}
\end{figure*}

In order to remove possible cognitive and cultural biases, we choose to design our experiment with abstract images randomly generated using two of the complexity measures presented above. The first set of images (Fig.~\ref{fig:images}a) is generated by  \textit{reverse-engineering} the Fourier Magnitude property: setting $\hat u(k,\theta) = k^\alpha e^{i2\pi \eta(\boldsymbol k)}$, where $\eta$ is drawn from a uniform distribution on $[0,1]$ with $\langle \eta(\boldsymbol k)\eta(\boldsymbol k')\rangle\propto \delta_{\boldsymbol k\boldsymbol k'}$, and taking the inverse Fourier Transform of $\hat u$ allows to produce a series of random greyscale images $|u(r,\phi)|$ with controlled Magnitude slope \cite{spehar2016taxonomy,Spehar2015,first_test} 
Table~\ref{tab:Fourier} gathers the computed complexity measures $\alpha$, $d_\mathrm{f}$ and $\tau$ of the 256$\times$256 images displayed in Fig.~\ref{fig:images}a. As one can see both $d_\mathrm{f}$ and $\tau$ are increasing functions of $\alpha$, comforting our choice of complexity measures and indicating that there is a clear correlation between the spectral, fractal and algorithmic properties.

The use of a second set of images was motivated by the remarks of some survey participants. When asked why they had preferred certain images, they responded their picks reminded them of cloudy skies or galactic landscapes. With the aim of producing more abstract images, we used an alternative method, now based on reverse-engineering the Minkowski-Bouligand box-counting method (Fig.~\ref{fig:images}b). Black squares of size $\epsilon = 2^{k}$, $k\in  [1,k_{\max}]$, drawn from the distribution $N_x(\epsilon)=A\epsilon^{-x}$ are randomly added to a white canvas of size 256$\times$256 with the condition that they do not overlap. The upper boundary  $k_{\max}$ is chosen such that the biggest squares occupy at most 1/16th of the total surface, $k_{\max}\leq\log_2(256)-2=6$. We also enforce that the total fraction of black pixels does not exceed 1/2. Here again the complexity measures appear to be increasing functions of one another (see Tab.~\ref{tab:Fourier}).

\begin{table}[b!]
\caption{\label{tab:Fourier}%
Complexity properties of the images presented in Fig.~\ref{fig:images}.}
\begin{ruledtabular}
\begin{tabular}{lcccccc}
&
\multicolumn{1}{c}{\textrm{a1}}&
\multicolumn{1}{c}{\textrm{a2}}&
\multicolumn{1}{c}{\textrm{a3}}&
\multicolumn{1}{c}{\textrm{a4}}&
\multicolumn{1}{c}{\textrm{a5}}&
\multicolumn{1}{c}{\textrm{a6}}\\
\hline
$\alpha$ &-3.96&-2.47&-1.95&-1.42&-0.76&0\\
d$_f$& 1.22 & 1.33 & 1.58 & 1.87 & 1.99 & 2.0 \\
$\tau$
  &0.045 & 0.074 & 0.12 & 0.22 & 0.40 & 0.41\\
\end{tabular}
\smallskip

\begin{tabular}{lcccccc}
&
\multicolumn{1}{c}{\textrm{b1}}&
\multicolumn{1}{c}{\textrm{b2}}&
\multicolumn{1}{c}{\textrm{b3}}&
\multicolumn{1}{c}{\textrm{b4}}&
\multicolumn{1}{c}{\textrm{b5}}&
\multicolumn{1}{c}{\textrm{b6}}\\
\hline
$\alpha$ &-2.14 &-1.78&-1.56&-1.32&-1.04&-0.77\\
d$_f$& 1.42 & 1.51 & 1.62 & 1.75 & 1.88 & 1.95 \\
$\tau$
  &0.012 & 0.014 & 0.022 & 0.044 & 0.095 & 0.16\\
\end{tabular}
\end{ruledtabular}
\end{table}

In 2013, Spehar and Taylor \cite{first_test} conducted a survey on twenty-six academics, using black and white computer generated images with increasing fractal dimension. They found a reversed U-shaped relation between image appreciation and Fractal dimension, with an aesthetic optimum for $d_\mathrm{f}\approx1.5$, allowing to argue that we indeed tend to prefer images with intermediate complexity, see also \cite{Spehar2015,spehar2016taxonomy}.
Curious of their results, we conducted a larger scale experiment intended for a larger panel (over a thousand participants with different backgrounds), using the images presented in Fig.~\ref{fig:images}. Our question at this stage is similar: is there a link between the statistical properties of our generated images and the tendency of people to appreciate them? 

Survey methods design constitutes a strand of research on its own \cite{LoshHesselbart1985}. For optimal results the selection task must be simple and display the minimum amount of information to the interviewee. While the common five-star ratings only take a time proportional to the number of images to score, these have been shown to be weighted by extreme grades, the utility given to intermediate grades being far from linear~\cite{aral2014problem}. Five-star ratings image-by-image can also be rather disorienting due to the lack of reference. Another option is image classification, where interviewee is presented with the whole set of images and is then asked to sort them by preference. While also time-efficient, presenting all the images at once might strongly induce people into intuitively recognising other features, such as complexity, and ending by sorting them by something else than preference. Finally, the \textit{Battle survey} consists in successively presenting to the interviewee all possible sets of two images and asking them to choose the one they prefer \cite{vessel2010beauty,mcmanus1980aesthetics}. While less time-efficient (with $N(N-1)/2=O(N^2)$ battles, for $N=$\,6 one needs 15 rounds to complete the survey), this method beats the other shortcomings mentioned above, and people usually feel more comfortable with such a binary task, intellectually less challenging. We thus choose the latter method. 

\begin{figure}[t!]
\includegraphics[width=0.9\columnwidth]{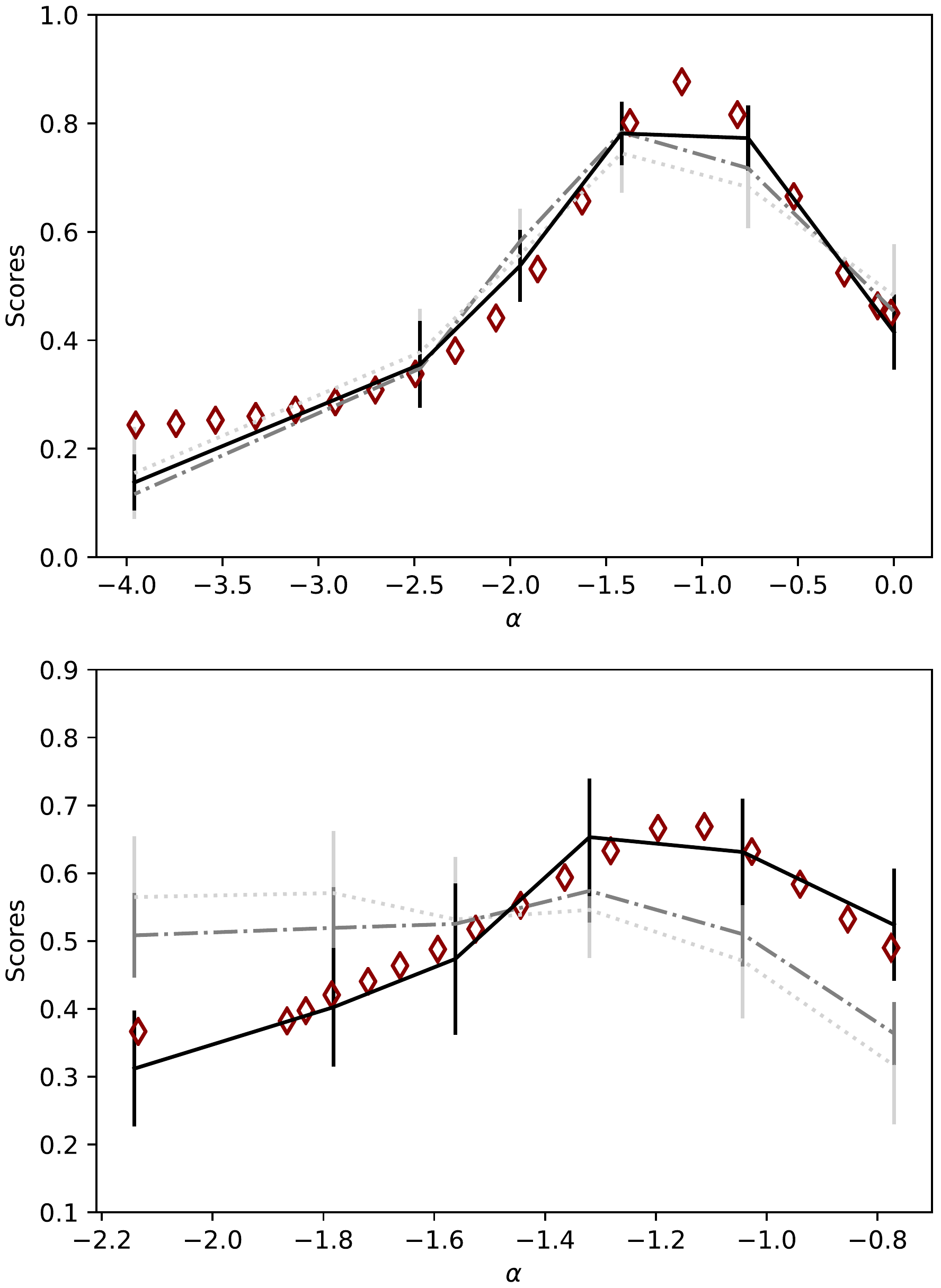}
\caption{\label{fig:scores} Results of the three different surveys (Zooniverse: solid black line, Mechanical Turk 1: dotted light gray-line,  Mechanical Turk 2: dash-dotted gray line). The diamonds markers indicate the structural complexity $\tau_\mathrm{cg}$ defined below. We have rescaled and shifted vertically $\tau_\mathrm{cg}$ to show that the maximum scores also correspond to maximum structural complexity. Top: image series of Fig.~\ref{fig:images}(a). Bottom: image series of  Fig.~\ref{fig:images}(b). Error bars reflect population heterogeneity.} 
\vspace{-0.2cm}
\end{figure}

We conducted three slightly different surveys. The panel for the first survey consisted of colleagues from CFM and Ecole Polytechnique as well students and relatives, adding up to $\approx 350$ people,  who were asked to participate without any financial incentive. While probably slightly biased population-wise, these are the results as we are most confident with, since we believe people in such a panel completed their tasks selflessly and honestly. To run this survey we used the Zooniverse platform~\cite{zoo} which provides a rather intuitive interface. The 15 two-image sets for each series were generated using a python algorithm that concatenated the images in a random order and attributed them a different name so that the interviewee couldn't find hidden information. Upon completion of the survey, to establish a global ranking of the images we attributed them a score according to the following rule: if image $i$ wins (resp. looses) a battle, its score $S_i$ increases (resp. decreases) by $1/N_i$ where $N_i$ denotes the number of battles in which $i$ was involved \footnote{Note that $N_i$ was not exactly equal to $N$ since a small number of participants stopped before completing the 15 battles.}. To obtain a score $S_i \in[0,1]$ we then transform it as $S_i\to (S_i+1)/2$. 
The results are plotted as a solid black line in Fig.~\ref{fig:scores}. Remarkably, the preferred images appear to be a4 a5, and b4 b5 respectively, both corresponding to $\alpha$ close to 1. To note, interestingly $\alpha \approx 1$ is often associated to the spectral properties of natural images \cite{stephens2013statistical,tolhurst1992amplitude}  and visual arts \cite{Koch2010}. Discussions with voters revealed that they found their favorite images to be the most \textit{harmonious} and \textit{well balanced}.

In order to increase the size and diversity of the panel, we ran two other experiments on the Mechanical Turk platform~\cite{mecturk}, in which participants are paid a small amount of money to participate (we reached $\approx 700$ panelists). The first attempt on this platform (dotted light-gray line in Fig.~\ref{fig:scores}) appeared to give rather noisy results, especially for the second set of images. We partly attribute this to a fraction of participants answering randomly to the battles, which is likely much less significant when people participate selflessly and in good will. In general, survey participants on such paying platforms are evaluated  and only get paid if they completed their tasks satisfactorily, but in our case one cannot really evaluate the panelists since \textit{satisfactorily} here only stands for \textit{honestly}. In the second tentative (dash-dotted gray line in Fig.~\ref{fig:scores}) we formulated the question differently in order to encourage non-random participation: much like in Keynes' famous beauty contest~\cite{Keynes}, people were asked to pick the image which they thought would be preferred by the majority and told they would not get paid if their overall choices fell too far off the average (needless to say, we did process all of the data, regardless of its distance to the average). While probably introducing other biases, the results displayed less noise and better agreement with that of the initial selfless survey. Also note that while all results are fully consistent for the first set of images (Fig. 1(a)), the second experiment leads to a less pronounced maximum for the image series of Fig. 1(b), with scores on average closer to $S=1/2$ than they were for the very first experiment.

Very much like entropy is used to measure the disorder in a physical system, we would now like to see whether there might exist a statistical proxy to estimate an image's \textit{harmony} and \textit{equilibrium}, as described by our survey participants. Given the complexity measures described above, images with low complexity display very simple shapes (a1 b1), and images with very high complexity display a large amount of white noise (a6 b6). Our survey revealed that maximum appreciation is obtained for intermediate complexity suggesting the following question: could it be that an aesthetically appealing image results from a subtle balance between complexity and regularity? And if so, can we find an associated statistical measure? The work of Desolneux \textit{et al.} \cite{Desolneux2008} clearly resonates with such questions. Guided by the idea that there is no perceptual structure in white noise, the authors attempted to characterise forms and structures and in particular defined unusual features or \textit{Gestalts} as \textit{sets of points whose (...) spatial arrangement could not occur in noise}. Their ideas can be easily illustrated with the coffee and cream dynamics \cite{coffee}. Consider the experiment in which plain cream is left to slowly mix with plain coffee. While the initial and final states of such a system display very regular homogeneous structures, the transitional regime displays interesting and \textit{complex} mixing patterns as the cream/coffee interface slowly disappears. So far we have used the term \textit{complexity} rather imprecisely and it is now time to  distinguish more rigorously two sorts of complexity. The first is \textit{entropic complexity} measuring the amount of information in the image, which in the coffee experiment can only be an increasing function of time according to the second law of thermodynamics; the second is \textit{structural complexity} accounting for the amount of features outside of
the noise, which here is a non-monotonous function of time displaying a maximum at intermediate stages where the non-trivial mixing patterns are most significant. Entropic complexity is well described by $\alpha$, $d_\mathrm{f}$ or more commonly $\tau$. Structural complexity, in a sense, measures \textit{noiseless} entropic complexity or \textit{interestingness}.

\begin{figure}[t!]
\includegraphics[width=1\columnwidth]{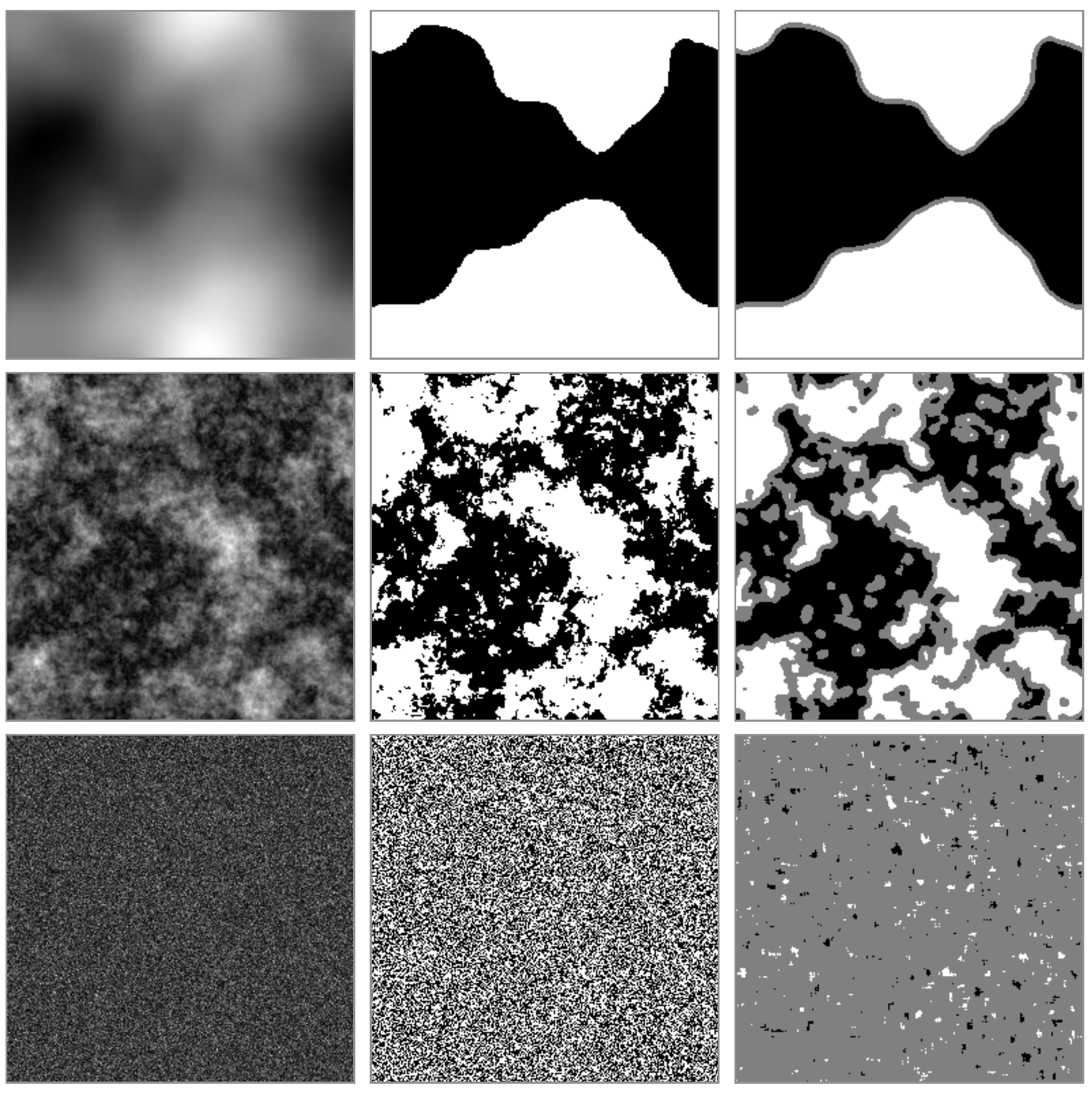}
\caption{\label{fig:blocking} Illustration of the coarse-graining procedure on images a1, a4 and a6, with $r_\mathrm{cg}=7$ and $\delta = 0.23$.}
\vspace{-0.2cm}
\end{figure}

Guided by the work of Aaronson \textit{et al.}~\cite{coffee} we computed structural complexity as a \textit{noiseless} entropy. 
More precisely we apply a coarse-graining procedure of given radius $r_\mathrm{cg}$ on the B\&W images and then compute their algorithmic complexity $\tau_\mathrm{cg}$ which we call structural complexity in the following \footnote{Spehar \textit{et al.} \cite{spehar2016taxonomy} showed that the preference curve was hardly affected by the gray scale to B\&W transformation. For the sake of simplicity we thus choose to apply the coarse-graining procedure to B\&W images.}. 
 The colour of a given block is determined by its black to white pixel ratio $\eta\in[0,1]$: white if $\eta\leq\delta$, gray if $\delta<\eta\leq1-\delta$, and black for $\eta>1-\delta$ where $\delta\in ]0,1/2[$ is a given threshold. Figure~\ref{fig:blocking} illustrates the procedure on images a1, a4 and a6; after turning them into B\&W (second column), the coarse-graining procedure is applied (third column). As one can see, image a1 is barely changed (just a thin gray line at the domain boundaries) and we thus expect $\tau_\mathrm{cg}\approx \tau$, image a4 is slightly denoised while letting its structures invariant $\tau_\mathrm{cg} \lesssim \tau$, image a6 however is strongly denoised as the coarse-graining procedure has left it almost plain gray suggesting $\tau_\mathrm{cg} \ll \tau$.  The structural complexity computed for both sets of image is plotted on Fig.~\ref{fig:scores} as dark red diamonds. As expected, $\tau_\mathrm{cg}(\alpha)$, or equivalently $\tau_\mathrm{cg}(d_\mathrm{f})$ and $\tau_\mathrm{cg}(\tau)$, are non-monotonous functions displaying a maximum for intermediate values of $\alpha$, $d_\mathrm{f}$ and $\tau$. 
Since the $y$-axis is rather arbitrary we have rescaled and shifted $\tau_\mathrm{cg}(\alpha)$ for an easier comparison with the survey data. 
Furthermore, the scale parameter $r_\mathrm{cg}$ and the threshold $\eta$ can be used as fitting parameters; in particular $r_\mathrm{cg}$ acts as the cutoff of a low-pass filter which erases high frequency spatial features, increasing it tends to lowers the right most red markers and shift the maximum to the left. Up to a multiplicative factor, best fits are obtained for $(r_\mathrm{cg},\eta)= (7,0.23)$ for the first set and $(13,0.12)$ for the second. 
The agreement between theory and experiments is quite convincing. Not only do the maxima coincide, but also the overall shape of the curves is similar. This quantitatively supports the idea that structural complexity is a good proxy for average image preference.

Let us summarise what we have achieved. Using two random-image-generation algorithms, we produced two different sets of abstract images spanning a broad range of entropic complexity, measured by three different quantities. We then designed and ran a large-scale experiment for image classification and found that preference peaks about complexity criteria matching that of natural images, perhaps indicating that people's preferences are influenced by their natural environment. Finally, our main contribution is to show that a ``noiseless'' entropy (that captures interesting structural features only) accounts well for the experimental results on image appreciation. It is interesting to speculate that, when confronted with images, the human brain may actually conduct the same kind of geometrical coarse-graining, trying to extract forms and structures while erasing uninteresting noise, or as put by the \textit{Gestalt theory} \cite{wertheimer1938gestalt}: filter meaningful perceptions from chaotic stimuli. As a result, the excess of noise and lack of forms may lead to unconscious rejection of
structureless images. 



\begin{acknowledgments}

We thank Christian Schmidt, Debanuj Chatterjee and Raphael Benichou for fruitful discussions, as well as Bastien Legay, who contributed to the early stages of this project. This publication uses data generated via the Zooniverse.org platform, development of which is funded by generous support, including a Global Impact Award from Google, and by a grant from the Alfred P. Sloan Foundation.
This research was conducted within the Econophysics \& Complex Systems Research Chair, under
the aegis of the Fondation du Risque, the Fondation de l'Ecole polytechnique, the Ecole polytechnique and Capital Fund Management.

\end{acknowledgments}

\nocite{*}

\bibliography{Lakhal}

\end{document}